\documentclass[12pt]{article}
\usepackage{xcolor}
\usepackage{cite}
\usepackage{qcircuit}
\usepackage{graphicx}
\usepackage[utf8]{inputenc}
\usepackage[dvips]{epsfig}
\usepackage{graphicx}
\usepackage[T1]{fontenc}
\usepackage{authblk}
\usepackage{amsmath}
\usepackage{float}
\usepackage{hyperref}
\date{}
\DeclareFixedFont{\ttb}{T1}{txtt}{bx}{n}{12} 
\DeclareFixedFont{\ttm}{T1}{txtt}{m}{n}{12}  

\usepackage{color}
\definecolor{deepblue}{rgb}{0,0,0.5}
\definecolor{deepred}{rgb}{0.6,0,0}
\definecolor{deepgreen}{rgb}{0,0.5,0}

\usepackage{listings}

\newcommand\pythonstyle{\lstset{
language=Python,
basicstyle=\ttm,
morekeywords={self},              
keywordstyle=\ttb\color{deepblue},
emph={MyClass,__init__},          
emphstyle=\ttb\color{deepred},    
stringstyle=\color{deepgreen},
frame=tb,                         
showstringspaces=false
}}

\lstnewenvironment{python}[1][]
{
\pythonstyle
\lstset{#1}
}
{}

\newcommand\pythoninline[1]{{\pythonstyle\lstinline!#1!}}
\title{\bf  Toward a quantum computing algorithm to quantify classical and quantum correlation of system states }
\author[1]{M. Mahdian\thanks{mahdian@tabrizu.ac.ir}}
\author[1]{H. Davoodi Yeganeh\thanks{h.yeganeh@tabrizu.ac.ir }}

\affil[1]{Faculty of Physics, Theoretical and astrophysics
department, University of Tabriz, 51665-163 Tabriz, Iran}

\begin{document}
\maketitle
\begin{abstract}

Optimal measurement is required to obtain the quantum and classical correlations of a quantum state, and the crucial difficulty is how to acquire the maximal information about one system by measuring the other part; in other words, getting the maximum information corresponds to preparing the best measurement operators. Within a general setup, we designed a variational hybrid quantum-classical (VHQC) algorithm to achieve classical and quantum correlations for system states under the Noisy-Intermediate Scale Quantum (NISQ)
technology. To employ, first, we map the density matrix to the vector representation, which displays it in a doubled Hilbert space, and it's converted to a pure state. Then we apply the measurement operators to a part of the subsystem and use variational principle and a classical optimization for the determination of the amount of correlation. We numerically test the performance of our algorithm at finding a correlation of some density matrices, and the output of our algorithm is compatible with the exact calculation.
\end{abstract}
\noindent
{\bf Keywords:Hybrid quantum-classical algorithm, Optimal measurement, Quantum correlation, Quantum discord}
\section{Introduction}

Quantum measurement is one of the most fundamental concepts of quantum mechanics that allows us to obtain the required information from the system \cite{r28}.  Hence, for gaining maximum information about the system, the best measurements (i.e., optimal measurements)  should be used. On the other hand,  finding the best measurements is one of the significant challenges in quantum computing.
Optimal measurements have been used in many context, such as quantum discrimination \cite{h4}, quantum entanglement\cite{h2,j1,j2}, quantum teleportation \cite{h1}, and superdense coding \cite{h3}, also in the calculation of quantum correlation\cite{r28,r29,j3,j4,j5,j6}.\\
We know that quantum correlations play a very critical role in quantum information and computation.  Quantities such as entanglement and quantum discord are used for measuring them. Obtaining and calculating correlations, whether quantum or classical, in physical systems, can help us to understand these systems more deeply. In computing correlations, measurements are performed on the system to gain correlation information and it is necessary again best measurements are employed \cite{r1,r5,j5,r3,r4,r6,r7,r8,r9,j7}.\\
Quantum entanglement, which is one of the most important features of quantum mechanics, and is widely used in quantum computing and information. It is often considered a criterion for quantum correlations. And based on that, quantum states can be divided into two categories; separable and entangled states.  It was introduced independently in \cite{r2,h5}, that entanglement is not the only type of quantum correlation, and there are separable states that have a quantum correlation. These quantum correlations, which are measured by quantum discord, may speed up some operations compared to their classical counterparts, so it's an important role in quantum information.\\
Quantum correlations and discord have received a great deal of attention in recent years, and numerous articles have been published on them \cite{h6,j8}. Quantum discord reduces to entanglement for pure states but has non-zero values for some mixed separable states. For mixed quantum states, it's defined as the difference in the total amount of correlation. Calculation of quantum discord is based on the minimization procedure on all possible measurements\cite{j9,j10,j11} on the subsystems, and thus, it is somewhat difficult to calculate even numerically.\\
Physical systems that can be represented mathematically by a density matrix and quantum correlations are encoded in this formulation. When the size of the system grows, it is complicated to determine the quantum correlation, in particular quantum discord, either analytically or computationally. Therefore, quantum discord has been calculated only for a somewhat limited set of two-qubit quantum states and is still an open problem for higher dimensions of quantum density matrices.
In the particular case, quantum discord derives explicit expressions for X-states and some bipartite quantum systems \cite{r4,r22,r31,luo} also an analytical solution obtained for quantum discord for $d\otimes 2$ systems\cite{r30} and have been investigated a class of two-qubit state with parallel nonzero Bloch vectors\cite{r23}.\\
Classical computers fail to efficiently simulate quantum systems with complex many-body
interactions due to the exponential growth of variables for characterizing these systems, so
the quantum simulation was proposed to solve such an exponential explosion problem using a
controllable quantum system \cite{r10,r11,r12}.
 Many quantum algorithms are used for quantum computing which will require many quantum resources and error correction, and due to the limiting their and technology today, usefulness soon \cite{r13}.\\
Among the different approaches to quantum computing, NISQ devices, which include relatively low-depth quantum circuits by hybrid variational quantum-classical algorithms, have recently received a lot of attention. Hybrid algorithms have been designed in a way that uses resources such as quantum and classical to solve problems specific optimization tasks that are not accessible to traditional classical computers \cite{r14,r15,r16,r17,r18,r19,x1,x4,x5,x6,x7,x8,x10,m11,m22}. One of the most important advantages of this method is that it requires a small number of qubits (contain from 10 to $10^3$ of qubits) to run with high gate fidelity and not fault-tolerant error correction \cite{r13}. Also, these algorithms have been introduced for other applications such as finding ground and excited states\cite{r19,r14,j12} and simulation system dynamics\cite{m11,m22,j13,j14}.\\
In this paper, we design the variational hybrid quantum-classical algorithm for calculating quantum and classical correlations. In this algorithm, using vector representation, the density matrix is mapped to a vector until performs unitary operations on pure states on NISQ devices.
In the first step, we directly encoding the quantum state $|\rho\rangle$ on the initial state $|0\rangle$ by using quantum circuits $U(x,y,z)$, which can include some rotations around axes $(x,y,z)$, $|\rho\rangle=U(x,y,z) |0\rangle.$
Then, we prepare the trial state $|\Psi(\vec{\theta})\rangle$ by by applying a sequence of parametrised measurement gates $R(\vec{\theta})$ on one of subsystems to quantum states $|\rho\rangle$. In the next step, by minimization entropy via gradient descent methods on the classical computer, the best measurements are obtainable to determine the classical and quantum correlation (details are given in section 3). We then numerically test the performance of our algorithm at finding classical and quantum correlation of some quantum states. The output of our algorithm is compatible with the results of analytical calculations.

The paper is structured as follows. In Sec.\ref{sec2}, we describe quantum discord.  In Sec.\ref{sec3}, we introduce the VHQC  algorithm that is used for calculation classical and quantum correlation. The performance of our algorithm to some density matrix will then be discussed in section 4, and Finally, Sec.\ref{sec5} gives the conclusions.
\section{Quantum discord}\label{sec2}

In classical information theory, for two sets of random variables $A$ and $B$ with the probabilities of occurrence $\{\mathbf{p}\}$ and $\{\mathbf{q}\}$ respectively, the mutual information represents the correlation of variables $A$ and $B$ and is written as follows:

\begin{equation}\label{m1}
I(A:B)= H(A)+H(B)-H(A,B),
\end{equation}
where, $H(A)=-\sum_i p(a_i)log ( p(a_i))$  and $H(B)=-\sum_i q(b_i)log (q(b_i)) $ are known as the Shannon entropy of the random
variables $A$ and $B$ and $p(x_i)$ is the probability of variable $x_i$. $H(A:B)=-\sum_{ij} p_{a_i,b_j}log\  p_{a_i,b_j}$ is joint Shannon entropy and $ p_{a_i,b_j}$ is the joint probability of variables $a_i$ and $b_j$ .
On the other hand, according to the definition of conditional entropy $H(A|B)=H(A,B)-H(B)$ which indicates the information gained about the subsystem $A$ by measuring the subsystem $B$, an alternative version of the mutual information can also be written as follows:
\begin{equation}\label{m2}
J(A:B)= H(A)-H(A|B).
\end{equation}
Note that, two expressions of mutual information, Eq.(\ref{m1}) and Eq.(\ref{m2})are equivalent in classic information theory but in the quantum regime, they can be different from each other. In quantum information, we use the density matrix, and von Neumann entropy instead of the  classical probability distributions and Shannon entropy, respectively, and the measure of the total correlations in a composite  system AB is given by the quantum mutual information as:
\begin{equation}
\mathcal{I}(\rho^{AB})=S(\rho^A)+S(\rho^B)-S(\rho^{AB}),
\end{equation}
where  $S(\rho)=-Tr(\rho log \rho)$ is the von Neumann entropy of state $\rho$. We suppose A and B share a quantum state $\rho^{AB}\in H_A\otimes H_B$ and $\rho^A = Tr_B(\rho^{AB})$ and $\rho^B = Tr_A(\rho^{AB})$ are the reduced density matrix of subsystem A and subsystem B, respectively.\\

Another definition of Equation Eq.(\ref{m2}) in quantum information is associated with the measurements we make on one of the subsystems.
\begin{equation}\label{E1}
\mathcal{J}(\rho^{AB}):= \{S(\rho^A)-S(\rho_{AB}|\{\Pi_B^j\})\}.
\end{equation}

Which, defines the maximum amount of information that can be obtained from subsystem A by measuring on subsystem B, and this definition is not symmetric. Also, $\{\Pi_B^j\}$ denotes a probability operator-valued measure (POVM) which describes a generalized measurement, and subscript $B$ indicates that the measurement is performed only on subsystem $B$.

\begin{equation}
S(\rho_{AB}|\{\Pi_B^j\}):=\sum_j p_j S(\rho_A|\{\Pi_B^j\})=\sum_j p_j S(\rho_j),
\end{equation}

and the reduced density matrix after the measurement is

\begin{equation}\label{q1}
\rho_j=\frac{1}{p_j}(I\otimes \Pi_B^j)\rho_{AB}(I\otimes \Pi_B^j),
\end{equation}
where $I$ is identity operator and
\begin{equation}\label{q2}
p_j=Tr[(I\otimes \Pi_B^j)\rho_{AB}(I\otimes \Pi_B^j)].
\end{equation}
which is the measurement probability for the jth projector.\\
Then the classical correlations are defined as the upper bound of $\mathcal{J}(\rho^{AB})$ and can be written according to the measurements on the subsystem $B$ as follows:
\begin{equation}\label{E2}
\mathcal{C}(\rho):= \underset{\{ \Pi_B^j\}}{sup}[ \mathcal{J}(\rho^{AB})]=S(\rho^A)- \underset{\{\Pi_B^j\}}{min}[S(\rho_{AB}|\{\Pi_B^j\})].
\end{equation}
which minimum is done on all possible measurements $\{ \Pi_B^j\}$.
Quantum discord is the difference of the total amount of correlation $\mathcal{I}(\rho)$ and the classical correlation defined by

\begin{equation}\label{E3}
\mathcal{QD}(\rho):=\mathcal{I}(\rho)-\mathcal{C}(\rho).
\end{equation}

\subsection{SU(N) algebra and density matrix}
As we know, any Hermitian operator and density matrix on N-dimensional Hilbert space can be written according to $SU(N)$ algebra generators.
Here, we consider $\lambda_i,\ i=1,2,...,N^2-1$  as the generators of $SU(N)$ that satisfy $Tr(\lambda_i \lambda_j)=2\delta_{ij}$ and $Tr(\lambda_i)=0$ \cite{su} which can be obtained by defining a set of $N$ projection operators as $P_{jk}=|j\rangle \langle k|$.
So we make $N^2-1$ operators as follows:

\begin{equation}\label{gen1}
U_{jk}=P_{jk}+P_{kj},
\end{equation}
\begin{equation}\label{gen2}
V_{jk}=-i(P_{jk}-P_{kj}),
\end{equation}
\begin{equation}\label{gen3}
W_l=\sqrt{\frac{2}{l(l+1)}}(P_{1l}+...+P_{ll}-l P_{l+1,l+1}), \quad 1\leq j < k \leq N  \quad  1\leq l\leq N-1
\end{equation}
and the set of the generators is given by
\begin{equation}\label{E22}
\{\lambda_i\}=\{U_{jk}\}\cup \{V_{jk}\}\cup \{W_l\},\quad  i=1,2,...,N^2-1
\end{equation}
By using this set of bases, the bipartite density matrix $\rho^{AB}$ on $mn$-dimensional  Hilbert $H^A_m \otimes H^B_n$ space can be  written as

\begin{equation}\label{d1}
\rho_{AB}=\frac{1}{mn}(I_m\otimes I_n +\sum_i \alpha_i \lambda_i\otimes I_n+\sum_jI \otimes  \beta_j \lambda_j +\sum_{i,j}m_{ij}\lambda_i\otimes \lambda_j),
\end{equation}
where $I$ is identity matrix and $\alpha_i$ , $ \beta_j$ and $m_{ij}$  are defined as
\begin{equation}
\alpha_i=\frac{1}{2n}Tr(\rho^{AB}\lambda_i\otimes I_n), \quad \beta_j=\frac{1}{2m}Tr(\rho^{AB}I_m\otimes\lambda_j ), \quad
m_{ij}=\frac{1}{4}Tr(\rho^{AB}\lambda_i\otimes \lambda_j).
\end{equation}
 Here, for example, we consider a two-qubit system state to evaluate quantum discord, therefore by using Eq.(\ref{d1}) a general two-qubit state can be written  as
 \begin{equation}\label{E33}
\rho_{AB}=\frac{1}{4}(I\otimes I +\sum_{i=3}^3 \alpha_i \sigma_i\otimes I +\sum_{j=1}^3 I \otimes  \beta_j \sigma_j +\sum_{i,j=1}^3m_{ij}\sigma_i\otimes \sigma_j),
\end{equation}
where, $\{\sigma_i\}_{i=3}^3$ are the Pauli matrices, $ \mathbf{\alpha}=(\alpha_1,\alpha_2,\alpha_3)$ and $ \mathbf{\beta}=(\beta_1,\beta_2,\beta_3)$ are  vector parameters of the subsystems A and B, respectively, and $m_{ij}$ are real numbers. By using of  singular value decomposition theorem, we can write matrix $M=\{m_{ij}\}$ as $M=UDV^T$, where U is an orthogonal matrix, $D$ is an diagonal matrix and $V$ is an orthogonal matrix. Moreover under
local unitary transformation of the form $(U_1\otimes U_2)\rho^{AB}(U_1^\dagger\otimes U_2^\dagger)$ with $U_1 , U_2 \in SU(2)$, and rewrite Eq(\ref{E33}) as
 \begin{equation}\label{E4}
\rho_{AB}=\frac{1}{4}(I\otimes I +\sum_{i=3}^3 \alpha_i \sigma_i\otimes I +\sum_{j=1}^3 I \otimes  \beta_j \sigma_j  +\sum_{j=1}^3 \omega_{j}\sigma_j\otimes \sigma_j).
\end{equation}
Which is the general form of the two qubits density matrix. For simplicity, we  consider the following simplified family of
states
 \begin{equation}\label{E55}
\rho_{AB}=\frac{1}{4}(I\otimes I +\sum_{j=1}^3 \omega_{j}\sigma_j\otimes \sigma_j).
\end{equation}
To evaluate, quantum discord $\mathcal{QD}(\rho)$ we need to evaluate $\mathcal{I}(\rho)$ and $\mathcal{C}(\rho)$.
After some calculation, we obtain
 \begin{equation}\label{E6}
\mathcal{I}(\rho)=2+\sum_{i=0}^3\nu_i log_2\nu_i,
\end{equation}
where $\nu_i$'s are eigenvalues of the density matrix . To  evaluate $\mathcal{C}(\rho)$, we need to perform the local measurements on part $B$ of the subsystem. By considering projection measurements as $\{|j\rangle \langle j|: j=0,1\}$, any von Neumann measurement can be written as $\{ \Pi_B^j=V |j\rangle \langle j| V^\dagger\ \}$. Here, $V \in SU(2)$  can be written as $V=rI+ i \vec{y}.\vec{\sigma}$ with $r\in \mathcal{R},\ \vec{y}=(y_1,y_2,y_3) \in \mathcal{R}^3$ and $ r^2+y_1^2+y_2^2+y_3^2=1$. After the measurement the density matrix of (\ref{E55}) transforme to the Eq.(\ref{q1}) and we have
\begin{equation}\label{param}
\begin{array}{ccl}
\rho_0=\frac{1}{p_0}(I\otimes \Pi_B^0).\frac{1}{4}(I\otimes I +\sum_{j=1}^3 \omega_{j}\sigma_j\otimes \sigma_j).(I\otimes \Pi_B^0)\\
\rho_1=\frac{1}{p_1}(I\otimes \Pi_B^1).\frac{1}{4}(I\otimes I +\sum_{j=1}^3 \omega_{j}\sigma_j\otimes \sigma_j).(I\otimes \Pi_B^1)\\
\end{array}
\end{equation}
and we have
\begin{equation}\label{param1}
\begin{array}{ccl}
\rho_0=\frac{1}{2}(I+\sum_{i=1}^3 \omega_i z_i\sigma_i)\otimes(V |0\rangle \langle 0|V^\dagger )\\
\rho_1=\frac{1}{2}(I-\sum_{i=1}^3 \omega_i z_i\sigma_i) \otimes(V|1\rangle \langle 1|V^\dagger),
\end{array}
\end{equation}
where,$z_1=2(-ry_2+y_1y_3),\ z_2=2(ry_2+y_2y_3),\ z_3=r^2+y^2_3-y_1^2-y_2^2$. So, we have;
\begin{equation}\label{s}
\begin{array}{ccl}
S(\rho_0)=S(\rho_1)=-\frac{1-\xi}{2}log_2\frac{1-\xi}{2} -\frac{1+\xi}{2}log_2\frac{1+\xi}{2}, \\
\end{array}
\end{equation}
and
\begin{equation}\label{ent1}
S(\rho_{AB}|\{\Pi_B^j\})=p_0 S(\rho_0)+p_1 S(\rho_1)=-\frac{1-\xi}{2}log_2\frac{1-\xi}{2} -\frac{1+\xi}{2}log_2\frac{1+\xi}{2},
\end{equation}
where $p_0=p_1=1/2$ and $\xi(r,y_1,y_2,y_3)=\sqrt{\sum_{i=1}^3|\omega_iz_i|^2}$. Using Eq.\ref{E2}, the classical correlation can be calculated as follows:
\begin{equation}\label{Eq2}
\mathcal{C}(\rho):= S(\rho^A)- \underset{\{\Pi_B^j(r,y_1,y_2,y_3)\}}{min}[-\frac{1-\xi}{2}log_2\frac{1-\xi}{2} -\frac{1+\xi}{2}log_2\frac{1+\xi}{2}],
\end{equation}
We see that $S(\rho_{AB}|\{\Pi_B^j\})$ depends on the set of von Neumann measurements and minimization is take over on all possible measurements or on a set of parameters $(r,y_1,y_2,y_3)$. \\
Therefore, for quantum discord, it must be possible to obtain the best measurement, which is described in terms of parameters, analytically or numerically. However, in this particular case for density matrix of \ref{E33}, these parameters can be found by analytical methods\cite{luo,r23,r4,r22}.

As can be seen, the most crucial step in calculating quantum discord is to obtain conditional entropy  $S(\rho_{AB}|\{\Pi_B^j\})$ and when the size of the system grows, it is complicated to determine. So, we introduce a new quantum algorithm to quantify classical and quantum correlation for calculating quantum correlations, which the details  will be explained in the next section.

\section{Quantum algorithm to quantify correlation}\label{sec3}

\begin{figure}[H]
\centering
\includegraphics[width=13.5cm]{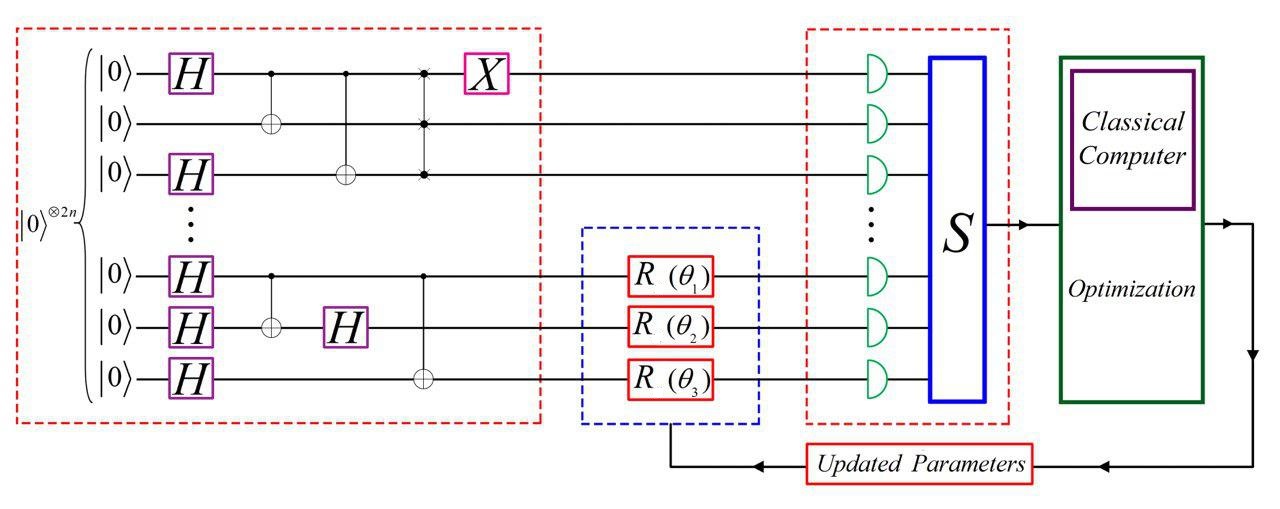}
\caption{A schematic of VHQC algorithm to quantify the quantum discord}\label{f}
\end{figure}

As mentioned, when the size of the system grows, it is complicated to quantify the quantum discord, either analytically or computationally.
Therefore, we introduce an algorithm VHQC, in which computations are divided between quantum and classical resources.
According to the equations  Eq.(\ref{E1}) $-$ Eq.(\ref{E3}), we need to minimize conditional entropy as a cost function. For this purpose, we represent the density matrix in the column form, $\rho\longrightarrow |\rho\rangle$ and $ \Pi \rho \Pi \longrightarrow \Pi \otimes \Pi^{T}|\rho\rangle$ and generate an trial state, $|\Psi(\vec{\theta})\rangle$, on a quantum processor by the action of a series of parameterized quantum measurements $R(\vec{\theta})$ and $\vec{\theta}=(\theta_1,\theta_2,...)$. The measurements are made by the Pauli matrices and then performed on $|\Psi(\vec{\theta})\rangle$ and the expected value of these matrices is calculated. Then, the conditional entropy with initial guess  parameter $\vec{\theta_0}$ calculated and fed to a classical minimization routine (e.g. gradient descent or Newton’s method) to indicate a new vector parameter $\vec{\theta_1}$.\\
A schematic of our algorithm is shown in Fig.[\ref{f}]. In the first layer, the initial state is prepared by using one and two-qubit gates. In the second layer, parametric measurement $R(\vec{\theta})$ are applied to one of the subsystems. It is then calculated entropy in the third layer after the measurement. Finally, in the last layer in a classical computer, we optimize the entropy-based on the previous results and propose new parameter values to improve the quantum measurement. By repeating these steps, the optimal parameter values are obtained to minimize the entropy.

For to preparation of the density matrix, we map the density matrix to a pure state using a doubled number of qubits as $|\rho\rangle \in \mathcal{H}\otimes  \mathcal{H}$.
\begin{equation}
\rho=\sum_{ij}\rho_{ij}|i\rangle \langle j|  \rightarrow |\rho\rangle =\sum_{ij}\frac{\rho_{ij}}{\sqrt{\sum_{ij}|\rho_{ij}|^2}} |i\rangle_p \otimes |j\rangle_A,
\end{equation}
where $|i\rangle_p$ and $|j\rangle_A$ are physical and ancillary qubits  respectively \cite{x5}.\\
In general form, we rewrite the unitary operators as a function of $\{\theta_i\}$
\begin{equation}\label{unit}
R(\vec{\theta})=R(\theta_1,\theta_2,...\theta_i....\theta_N)=R(\theta_1)R(\theta_2)...R(\theta_i)...R(\theta_N),\\
\end{equation}
Each of the unitary measurement operators $R(\theta_i)$ is dependent on only one parameter and in terms of Hermitian operators $\{\Lambda_i\}$:
\begin{equation}\label{unit22}
R(\theta_i)=exp(-i \theta_i \Lambda_i)=exp(-i\sum_{j} \theta_i s_{i,j} \hat{\sigma}_{i,j}),
\end{equation}
and $\Lambda_i=\sum_{j} s_{i,j} \hat{\sigma}_{i,j}$ where $\hat{\sigma}_{i,j}$ are Pauli operators.
The measurements that we consider in this article are of the Von Neumann measurement type, which includes a set of one-dimensional projection operators with sum up to the identity $\sum_{j} \Pi_B^j=I$ . Projection operators $\{\Pi_B^j\}$ describe a von Neumann measurement for one of the subsystems only. In general, operators $\{\Pi_B^j\}$ may be defined as a variational form
\begin{equation}\label{E77}
\Pi_B^j\equiv R(\vec{\theta_j})
\end{equation}
In the next step, a von Neumann measurement applies to one of the subsystems only that is defined in Eq.(\ref{E77}). These parameter gates can be implemented using single-qubit gates in the quantum circuit.
 These two-step implement in a quantum computer, based on the measurement results in a quantum computer, the classical computer computes new parameters by using optimization methods and sent to the quantum computer. The last three steps are repeated until quantity $S(\rho_{AB}|\{\Pi_B^j\})$ converges.


\subsection{Optimization via Gradient Descent}
Within the context of hybrid quantum-classical algorithms,
gradient descent optimizers require measuring the expectation values of parameterized quantities in quantum circuits.
Here to use a gradient-based approach for our algorithm, we can write the cost function as:
\begin{equation}
S(\rho_{AB}|\{\Pi_B^j (\vec{\theta})\}):=\sum_j p_j S(\rho_j(\vec{\theta}),
\end{equation}
where $\vec{\theta}$ is variational parameters and the partial derivative of this cost function with respect
to $\vec{\theta}$ is
\begin{equation}\label{p1}
\frac{\partial S(\rho_{AB}|\{\Pi_B^j (\vec{\theta})\})}{\partial \vec{\theta}}=\sum_j p_j \{-Tr(\frac{\partial \rho_j(\vec{\theta})}{\partial \vec{\theta}}log_2\rho_j(\vec{\theta})+\frac{\partial \rho_j(\vec{\theta})}{\partial \vec{\theta}})\}.
\end{equation}
Therefore, using Eq.(\ref{p1}) for parameters $y_k$  $(k=1,2,3)$ and $r$, we can evaluate the gradient descent of $S(\rho_{AB}|\{\Pi_B^j (\vec{\theta})\})$
directly and use a iteration as follows
$$\theta_k^{(t+1)}=\theta_k^{(t)}-\eta \sum_j p_j \{-Tr(\frac{\partial \rho_j}{\partial \theta_k}log_2\rho_j+\frac{\partial \rho_j}{\partial \theta_k})\},  $$ to minimize the cost function.

Let us now outline the main steps in our algorithm to calculate quantum discord:\\
(i) we prepare the density matrix $|\rho\rangle$ on the quantum computer.\\
(ii)Then, we apply the von Neumann measurement to one of the subsystems according to initial parameters, and then measure the entropy values.\\
(iii) we use the gradient-descent based method to optimize parameters and then determine the new values of parameters.\\
(iv) Iterate this procedure until convergence in the value of the parameters and find the best measurement based on minimum entropy.

\section{Numerical examples}\label{sec4}
\begin{figure}[H]
\centering
\includegraphics[width=13.5cm]{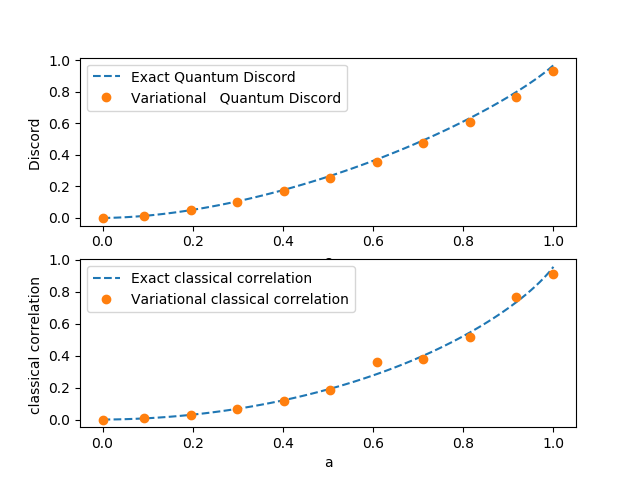}
\caption{Quantum discord and classical correlation by exact(blue dash line) values and our algorithm(orange dot) over a range of parameters $\{a\}$ for Werner state Eq.(\ref{E5}). The error of our method is less than  $ 5\times 10^{-6}$ for all parameters $\{a\}$.   }\label{f2}
\end{figure}

We simulate our  algorithm on the family of
states $\rho=\frac{1}{4}(I+\sum_{j=1}^3 c_j\sigma_j\otimes \sigma_j)$,  where $c_j$ are real constants and $\sigma_j$ are Pauli matrix's. These state's in general form called X-state and quantum discord derive explicitly
expressions for them \cite{r4,r22}. In the first example, we consider the Werner state \cite{r25}
\begin{equation}\label{E5}
\rho=a |\psi^-\rangle \langle \psi^-| +\frac{1-a}{4}I
\end{equation}

where $|\psi^-\rangle = \frac{1}{\sqrt{2}}(|01\rangle -|10\rangle)$ is an entangled
bell state and $0 \leq a \leq 1$. The output of the VHQC algorithm compared it to exact quantum discord and classical correlation as shown in Fig.[\ref{f2}].

\begin{figure}[H]
\centering
\includegraphics[width=13.5cm]{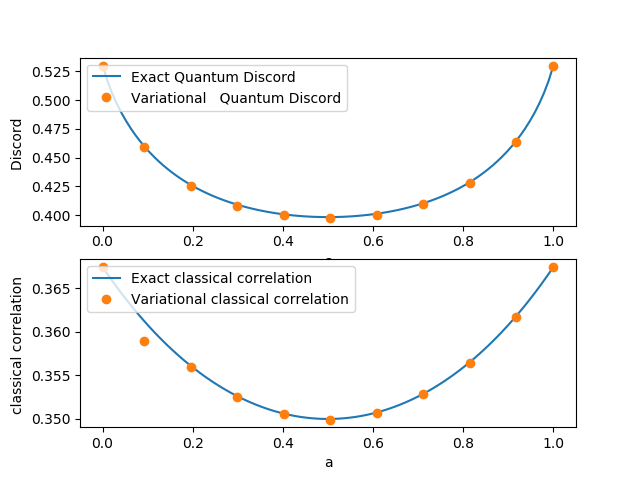}
\caption{Quantum discord and classical correlation by exact(blue line) values and our algorithm(orange dot) over a range of parameters $\{a\}$ for state   Eq.(\ref{E6}). Error of our method is less than  $ \approx10^{-8}$ for all  parameters $\{a\}$.}\label{f3}
\end{figure}

\begin{figure}[H]
\centerline{
\Qcircuit @C=1em @R=0.5cm { \lstick{|0\rangle} & \qw  &\gate{H} & \ctrl{1} & \qw &\qw&\targ &\qw &\qw &\qw & \multigate{3}{M}\qw &\qw &   \multigate{3}{S}&\multigate{3}{O}\qw &\qw
 \\ \lstick{|0\rangle} & \qw &\qw &  \targ & \qw &  \targ &\qw &\qw &\qw &\qw &\ghost{M}&\qw &\ghost{S}&\ghost{O}&\qw &
 \\ \lstick{|0\rangle}  &  \qw  &\gate{H} & \ctrl{1}  & \gate{H}& \qw& \ctrl{-2}& \qw & \multigate{1}{U} &\qw &\ghost{M}&\qw &\ghost{S}&\ghost{O}&\qw &
 \\ \lstick{|0\rangle}  &  \qw &\qw &  \targ &  \qw & \ctrl{-2}  &\qw &\qw &\ghost{U}&\qw &\ghost{M}&\qw&\ghost{S}&\ghost{O}&\qw
\gategroup{1}{2}{4}{13}{.7em}{--}
}
}
\caption{A schematic of our VHQC that applied for examples. U stands for general operators $\{B_j\}$, M stands for measurement, S stands for entropy, and O is used for the optimization of parameters in classical computers. Operations inside the dashed box are implemented in quantum computers. }\label{f4}
\end{figure}
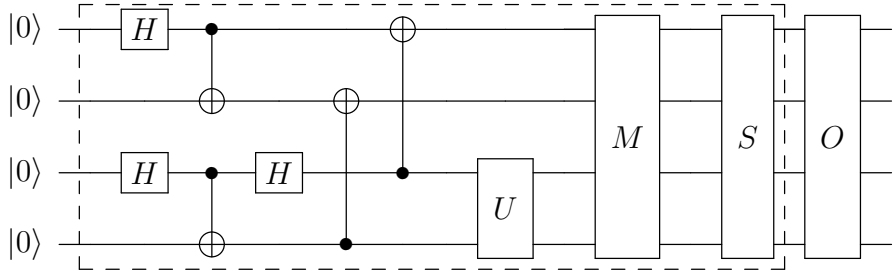

In the second example we consider the state
\begin{equation}\label{E6}
\rho=\frac{1}{3}[(1-a)|00\rangle \langle 00| + |\psi^+\rangle \langle \psi^+| + a|11\rangle \langle 11|]
\end{equation}
where $|\psi^+\rangle = \frac{1}{\sqrt{2}}(|01\rangle+|10\rangle)$ is an  entangled
bell state and $0 < a \leq 1$. Similar to the previous example output of VHQC algorithm compared it to exact quantum discord and classical correlation as shown in Fig.[\ref{f3}].
Also, we propose a schematic of our VHQC algorithm used in the previous two examples shown in Fig.[\ref{f4}]. Dash box connecting to employ hybrid algorithm on a quantum computer.
In the last example let us consider a random state discussed in Ref.\cite{r26}
\begin{equation}
\rho=
\begin{pmatrix}
0.437&0.126+0.197i&0.0271-0.0258i&-0.247+0.0997i \\ 0.126-0.197i&0.154&-0.0115-0.0187i&-0.0315+0.170i \\ 0.0271+0.0258i&-0.0115+0.0187i&0.0370&0.00219-0.0367i \\ -0.247-0.0997i&-0.0315-0.170i&0.00219+0.0367i&0.372
\end{pmatrix}
\end{equation}
we then perform a projective measurement on subsystem B and using the VHQC algorithm, the minimum of conditional entropy was obtained as 0.24 which is compatible with the value obtained for conditional entropy in Ref.\cite{r23}.
\section{Conclusions}\label{sec5}

We have proposed a VHQC quantum algorithm under NISQ devices to find the best measurement operator and then calculated the classical and correlations of a quantum system. We used the density matrix-vector to prepare the states, and then we were able to measure the quantum discord using the classical optimization method. We then numerically have tested the performance of our algorithm at finding quantum correlations of families of some quantum mixed states. The results of our algorithm have compatible with the exact calculation.

\newpage
{\bf Appendix I}\\
\noindent
Python code for  measuring the expectation
values of parameterized quantities in quantum circuit.
\begin{python}
from qutip import *
import numpy as np
from numpy import*
from math import*
from random import*
from scipy.optimize import*
from scipy import optimize
import random

"""
Within the context of hybrid quantum-classical algorithms,
gradient descent optimizers require measuring the expectation
values of parameterized quantities in quantum circuits.
Here to used a gradient-based approach for our algorithm.
"""

"""
define basis of two qubit system
"""
ket0=basis(2,0)
ket1=basis(2,1)
ket00=tensor(ket0,ket0);ket01=tensor(ket0,ket1);
ket10=tensor(ket1,ket0);ket11=tensor(ket1,ket1);

angle = np.linspace(0.0, 2 * np.pi, 100)


"""
define von Neumann measurement
"""
def R(theta,n_x,n_y,n_z):
    r=(cos(theta/2)-1j*n_z*sin(theta/2))*ket0*ket0.dag()
    -(1j*(n_x-1j*n_y)*sin(theta/2))*ket0*ket1.dag()
    -(1j*(n_x+1j*n_y)*sin(theta/2))*ket1*ket0.dag()
    (cos(theta/2)+1j*n_z*sin(theta/2))*ket1*ket1.dag()
    return r

def B1(theta,n_x,n_y,n_z):
    b1=(cos(theta/2)**2+ n_z**2*sin(theta/2)**2)*ket0*ket0.dag()
    +(1j*(n_x-1j*n_y)*sin(theta/2)*(cos(theta/2)
    -1j*n_z*sin(theta/2)))*ket0*ket1.dag()
    -(1j*(n_x+1j*n_y)*sin(theta/2)*(cos(theta/2)
    +1j*n_z*sin(theta/2)))*ket1*ket0.dag()
    +((n_x**2+ n_y**2)*sin(theta/2)**2)*ket1*ket1.dag()
    return b1

def B2(theta,n_x,n_y,n_z):
    b2=(cos(theta/2)**2+ n_z**2*sin(theta/2)**2)*ket1*ket1.dag()
        -(1j*(n_x-1j*n_y)*sin(theta/2)*(cos(theta/2)
       -1j*n_z*sin(theta/2)))*ket0*ket1.dag()
       +(1j*(n_x+1j*n_y)*sin(theta/2)*(cos(theta/2)
       +1j*n_z*sin(theta/2)))*ket1*ket0.dag()
       +((n_x**2+ n_y**2)*sin(theta/2)**2)*ket0*ket0.dag()
    return b2

"""
define density matrix, Here for simplicity, we set a=1
"""

rho=0.25*ket00*ket00.dag()+0.25*ket00*ket11.dag()
       +0.25*ket01*ket01.dag() +0.25*ket01*ket10.dag()
      +0.25*ket10*ket01.dag()+0.25*ket10*ket10.dag()
      +0.25*ket11*ket00.dag()+0.25*ket11*ket11.dag()

"""
parametric states"""

def P1(theta,n_x,n_y,n_z):
    p1=1/(tensor(qeye(2),B1(theta,n_x,n_y,n_z))).tr()*tensor(qeye(2)
    ,B1(theta,n_x,n_y,n_z))*rho*tensor(qeye(2),B1(theta,n_x,n_y,n_z))
    return p1

def P2(theta,n_x,n_y,n_z):
    p2=1/(tensor(qeye(2),B2(theta,n_x,n_y,n_z))).tr()*tensor(qeye(2)
    ,B2(theta,n_x,n_y,n_z))*rho*tensor(qeye(2),B2(theta,n_x,n_y,n_z))
    return p2


'define conditional entropy'
def S(theta,n_x,n_y,n_z):
    p1=1/(tensor(qeye(2),B1(theta,n_x,n_y,n_z))).tr()*tensor(qeye(2),
    B1(theta,n_x,n_y,n_z))*rho*tensor(qeye(2),B1(theta,n_x,n_y,n_z))

    p2=1/(tensor(qeye(2),B2(theta,n_x,n_y,n_z))).tr()*tensor(qeye(2)
   ,B2(theta,n_x,n_y,n_z))*rho*tensor(qeye(2),B2(theta,n_x,n_y,n_z))

    s=1/(tensor(qeye(2),B1(theta,n_x,n_y,n_z))).tr()*entropy_vn(p1)
      +1/(tensor(qeye(2),B2(theta,n_x,n_y,n_z))).tr()*entropy_vn(p2)
    return s

'Optimization step'
list3=[]
list4=[]
list5=[]

for i in range(100):
    t=uniform(0,1)
    y=uniform(0,1)
    z=uniform(0,1)
    if t**2+y**2+z**2==1:
       continue
    list3.append(t)
    list4.append(y)
    list5.append(z)

for theta,n_x,n_y,n_z in zip(angle,list3,list4,list5):
    initial_guess=[0,0,0,0]
    res = minimize(S,initial_guess ,method='Nelder-Mead',
                         options={'xtol': 1e-8, 'disp': True})
    print(res.x)









\end{python}

\bibliographystyle{ieeetr}
\bibliography{ref}

\begin{thebibliography}{10}

\bibitem{r28}
M.~A. Nielsen and I.~L. Chuang, ``Quantum computation and quantum
  information,'' {\em Phys. Today}, vol.~54, pp.~60--2, 2001.

\bibitem{h4}
J.~A. Bergou, ``Quantum state discrimination and selected applications,'' in
  {\em Journal of Physics: Conference Series}, vol.~84, p.~012001, IOP
  Publishing, 2007.

\bibitem{h2}
N.~Milazzo, D.~Braun, and O.~Giraud, ``Optimal measurement strategies for fast
  entanglement detection,'' {\em Physical Review A}, vol.~100, no.~1,
  p.~012328, 2019.

\bibitem{j1}
M.~Jafarizadeh, Y.~Akbari, K.~Aghayar, A.~Heshmati, and M.~Mahdian,
  ``Investigating a class of 2$\otimes $2 $\otimes $ d bound entangled density
  matrices via linear and nonlinear entanglement witnesses constructed by exact
  convex optimization,'' {\em Physical Review A}, vol.~78, no.~3, p.~032313,
  2008.

\bibitem{j2}
M.~Jafarizadeh, M.~Mahdian, A.~Heshmati, and K.~Aghayar, ``Detecting some
  three-qubit mub diagonal entangled states via nonlinear optimal entanglement
  witnesses,'' {\em The European Physical Journal D}, vol.~50, no.~1,
  pp.~107--121, 2008.

\bibitem{h1}
S.~Harraz, J.~Yang, K.~Li, and S.~Cong, ``Quantum state transfer control based
  on the optimal measurement,'' {\em Optimal Control Applications and Methods},
  vol.~38, no.~5, pp.~744--753, 2017.

\bibitem{h3}
A.~Abeyesinghe, P.~Hayden, G.~Smith, and A.~J. Winter, ``Optimal superdense
  coding of entangled states,'' {\em IEEE transactions on information theory},
  vol.~52, no.~8, pp.~3635--3641, 2006.

\bibitem{r29}
K.~Modi, A.~Brodutch, H.~Cable, T.~Paterek, and V.~Vedral, ``Quantum discord
  and other measures of quantum correlation,'' {\em Rev. Mod. Phys}, vol.~84,
  p.~1655, 2012.

\bibitem{j3}
M.~Mahdian, R.~Yousefjani, and S.~Salimi, ``Quantum discord evolution of
  three-qubit states under noisy channels,'' {\em The European Physical Journal
  D}, vol.~66, no.~5, p.~133, 2012.

\bibitem{j4}
M.~Mahdian, M.~Jeddi, M.~Yahyavi, and M.~Marahem, ``Dynamics of quantum
  dissonance of two qubit xxz model with dzyloshinsky-moriya interaction
  coupled to non-markovian environment,'' {\em International Journal of
  Theoretical Physics}, vol.~52, no.~11, pp.~3830--3843, 2013.

\bibitem{j5}
M.~Mahdian, M.~Yahyavi, and R.~Yousefjani, ``Correlation dynamics of
  three-qubit system under a classical dephasing environment,'' {\em
  International Journal of Theoretical Physics}, vol.~53, no.~1, pp.~203--215,
  2014.

\bibitem{j6}
M.~Mahdian, B.~Mojaveri, A.~Dehghani, and T.~Makaremi, ``Quantum correlations
  of two relativistic spin-$ \frac{1}{2} $ particles under noisy channels,''
  {\em International Journal of Theoretical Physics}, vol.~55, no.~2,
  pp.~678--697, 2016.

\bibitem{r1}
F.~Galve, G.~L. Giorgi, and R.~Zambrini, ``Quantum correlations and
  synchronization measures,'' in {\em Lectures on general quantum correlations
  and their applications}, pp.~393--420, Springer, 2017.

\bibitem{r5}
M.~Qin, Z.-Z. Ren, and X.~Zhang, ``Renormalization of the global quantum
  correlation and monogamy relation in the anisotropic heisenberg xxz model,''
  {\em Quantum Information Processing}, vol.~15, no.~1, pp.~255--267, 2016.

\bibitem{r3}
M.~Allegra, P.~Giorda, and A.~Montorsi, ``Quantum discord and classical
  correlations in the bond-charge hubbard model: Quantum phase transitions,
  off-diagonal long-range order, and violation of the monogamy property for
  discord,'' {\em Physical Review B}, vol.~84, no.~24, p.~245133, 2011.

\bibitem{r4}
L.~Qiu, G.~Tang, X.-q. Yang, and A.-m. Wang, ``Relating tripartite quantum
  discord with multisite entanglement and their performance in the
  one-dimensional anisotropic xxz model,'' {\em EPL (Europhysics Letters)},
  vol.~105, no.~3, p.~30005, 2014.

\bibitem{r6}
N.~Lambert, Y.-N. Chen, Y.-C. Cheng, C.-M. Li, G.-Y. Chen, and F.~Nori,
  ``Quantum biology,'' {\em Nature Physics}, vol.~9, no.~1, p.~10, 2013.

\bibitem{r7}
K.~R.~K. Rao, H.~Katiyar, T.~Mahesh, A.~Sen, U.~Sen, A.~Kumar, {\em et~al.},
  ``Multipartite quantum correlations reveal frustration in a quantum ising
  spin system,'' {\em Physical Review A}, vol.~88, no.~2, p.~022312, 2013.

\bibitem{r8}
S.~Rodriques, B.~Brock, P.~Love, J.~Zhu, S.~Kais, and A.~Aspuru-Guzik,
  ``Multipartite quantum entanglement evolution in photosynthetic complexes,''
  in {\em APS April Meeting Abstracts}, 2012.

\bibitem{r9}
T.~Chanda, U.~Mishra, A.~S. De, and U.~Sen, ``Time dynamics of multiparty
  quantum correlations indicate energy transfer route in light-harvesting
  complexes,'' {\em arXiv preprint arXiv:1412.6519}, 2014.

\bibitem{j7}
M.~Mahdian and H.~Kouhestani, ``Thermal quantum correlations in photosynthetic
  light-harvesting complexes,'' {\em International Journal of Theoretical
  Physics}, vol.~54, no.~8, pp.~2576--2590, 2015.

\bibitem{r2}
H.~Ollivier and W.~H. Zurek, ``Quantum discord: a measure of the quantumness of
  correlations,'' {\em Physical review letters}, vol.~88, no.~1, p.~017901,
  2001.

\bibitem{h5}
L.~Henderson and V.~Vedral, ``Classical, quantum and total correlations,'' {\em
  Journal of physics A: mathematical and general}, vol.~34, no.~35, p.~6899,
  2001.

\bibitem{h6}
A.~Bera, T.~Das, D.~Sadhukhan, S.~S. Roy, A.~S. De, and U.~Sen, ``Quantum
  discord and its allies: a review of recent progress,'' {\em Reports on
  Progress in Physics}, vol.~81, no.~2, p.~024001, 2017.

\bibitem{j8}
M.~Mahdian and M.~B. Arjmandi, ``Comparison of quantum discord and relative
  entropy in some bipartite quantum systems,'' {\em Quantum Information
  Processing}, vol.~15, no.~4, pp.~1569--1583, 2016.

\bibitem{j9}
E.~Mart{\'\i}nez-Vargas, C.~Pineda, and P.~Barberis-Blostein, ``Quantum
  measurement optimization by decomposition of measurements into extremals,''
  {\em Scientific Reports}, vol.~10, no.~1, pp.~1--10, 2020.

\bibitem{j10}
L.~A. de~Castro, O.~P. d.~S. Neto, and C.~A. Brasil, ``An introduction to
  quantum measurements with a historical motivation,'' {\em arXiv preprint
  arXiv:1908.03949}, 2019.

\bibitem{j11}
M.~Burgos, ``The measurement problem in quantum mechanics revisited,'' {\em
  Selected Topics in Applications of Quantum Mechanics, INTECH, Croatia},
  pp.~137--173, 2015.

\bibitem{r22}
M.~Ali, A.~Rau, and G.~Alber, ``Quantum discord for two-qubit x states,'' {\em
  Physical Review A}, vol.~81, no.~4, p.~042105, 2010.

\bibitem{r31}
A.~Rau, ``Calculation of quantum discord in higher dimensions for x-and other
  specialized states,'' {\em Quantum Information Processing}, vol.~17, no.~9,
  p.~216, 2018.

\bibitem{luo}
S.~Luo, ``Quantum discord for two-qubit systems,'' {\em Physical Review A},
  vol.~77, no.~4, p.~042303, 2008.

\bibitem{r30}
Z.~Ma, Z.~Chen, F.~F. Fanchini, and S.-M. Fei, ``Quantum discord for $ d\otimes
  2 $systems,'' {\em Scientific reports}, vol.~5, p.~10262, 2015.

\bibitem{r23}
D.~Girolami and G.~Adesso, ``Quantum discord for general two-qubit states:
  analytical progress,'' {\em Physical Review A}, vol.~83, no.~5, p.~052108,
  2011.

\bibitem{r10}
S.~Lloyd, M.~Mohseni, and P.~Rebentrost, ``Quantum principal component
  analysis,'' {\em Nature Physics}, vol.~10, no.~9, p.~631, 2014.

\bibitem{r11}
S.~Garnerone, P.~Zanardi, and D.~A. Lidar, ``Adiabatic quantum algorithm for
  search engine ranking,'' {\em Physical review letters}, vol.~108, no.~23,
  p.~230506, 2012.

\bibitem{r12}
A.~Aspuru-Guzik, A.~D. Dutoi, P.~J. Love, and M.~Head-Gordon, ``Simulated
  quantum computation of molecular energies,'' {\em Science}, vol.~309,
  no.~5741, pp.~1704--1707, 2005.

\bibitem{r13}
J.~Preskill, ``Quantum computing in the nisq era and beyond,'' {\em Quantum},
  vol.~2, p.~79, 2018.

\bibitem{r14}
J.~R. McClean, J.~Romero, R.~Babbush, and A.~Aspuru-Guzik, ``The theory of
  variational hybrid quantum-classical algorithms,'' {\em New Journal of
  Physics}, vol.~18, no.~2, p.~023023, 2016.

\bibitem{r15}
E.~Farhi, J.~Goldstone, and S.~Gutmann, ``A quantum approximate optimization
  algorithm,'' {\em arXiv preprint arXiv:1411.4028}, 2014.

\bibitem{r16}
D.~Wang, O.~Higgott, and S.~Brierley, ``A generalised variational quantum
  eigensolver,'' {\em arXiv preprint arXiv:1802.00171}, 2018.

\bibitem{r17}
P.~D. Johnson, J.~Romero, J.~Olson, Y.~Cao, and A.~Aspuru-Guzik, ``Qvector: an
  algorithm for device-tailored quantum error correction,'' {\em arXiv preprint
  arXiv:1711.02249}, 2017.

\bibitem{r18}
S.~Endo, T.~Jones, S.~McArdle, X.~Yuan, and S.~Benjamin, ``Variational quantum
  algorithms for discovering hamiltonian spectra,'' {\em arXiv preprint
  arXiv:1806.05707}, 2018.

\bibitem{r19}
A.~Peruzzo, J.~McClean, P.~Shadbolt, M.-H. Yung, X.-Q. Zhou, P.~J. Love,
  A.~Aspuru-Guzik, and J.~L. O’brien, ``A variational eigenvalue solver on a
  photonic quantum processor,'' {\em Nature communications}, vol.~5, p.~4213,
  2014.

\bibitem{x1}
R.~LaRose, A.~Tikku, {\'E}.~O’Neel-Judy, L.~Cincio, and P.~J. Coles,
  ``Variational quantum state diagonalization,'' {\em npj Quantum Information},
  vol.~5, no.~1, pp.~1--10, 2019.

\bibitem{x4}
A.~N. Chowdhury, G.~H. Low, and N.~Wiebe, ``A variational quantum algorithm for
  preparing quantum gibbs states,'' {\em arXiv preprint arXiv:2002.00055},
  2020.

\bibitem{x5}
N.~Yoshioka, Y.~O. Nakagawa, K.~Mitarai, and K.~Fujii, ``Variational quantum
  algorithm for non-equilirium steady states,'' {\em arXiv preprint
  arXiv:1908.09836}, 2019.

\bibitem{x6}
M.~Lubasch, J.~Joo, P.~Moinier, M.~Kiffner, and D.~Jaksch, ``Variational
  quantum algorithms for nonlinear problems,'' {\em Physical Review A},
  vol.~101, no.~1, p.~010301, 2020.

\bibitem{x7}
S.~Endo, Y.~Li, S.~Benjamin, and X.~Yuan, ``Variational quantum simulation of
  general processes,'' {\em arXiv preprint arXiv:1812.08778}, 2018.

\bibitem{x8}
S.~McArdle, T.~Jones, S.~Endo, Y.~Li, S.~C. Benjamin, and X.~Yuan,
  ``Variational ansatz-based quantum simulation of imaginary time evolution,''
  {\em npj Quantum Information}, vol.~5, no.~1, pp.~1--6, 2019.

\bibitem{x10}
M.~Cerezo, K.~Sharma, A.~Arrasmith, and P.~J. Coles, ``Variational quantum
  state eigensolver,'' {\em arXiv preprint arXiv:2004.01372}, 2020.

\bibitem{m11}
M.~Mahdian and H.~Davoodi~Yeganeh, ``Hybrid quantum variational algorithm for
  simulating open quantum systems with near-term devices,'' {\em Journal of
  Physics A: Mathematical and Theoretical}, vol.~53, no.~41, p.~415301, 2020.

\bibitem{m22}
M.~Mahdian and H.~Davoodi~Yeganeh, ``Incoherent quantum algorithm dynamics of
  an open system with near-term devices,'' {\em Quantum Information
  Processing}, vol.~19, no.~9, pp.~1--13, 2020.

\bibitem{j12}
K.~M. Nakanishi, K.~Mitarai, and K.~Fujii, ``Subspace-search variational
  quantum eigensolver for excited states,'' {\em Physical Review Research},
  vol.~1, no.~3, p.~033062, 2019.

\bibitem{j13}
S.~Endo, J.~Sun, Y.~Li, S.~C. Benjamin, and X.~Yuan, ``Variational quantum
  simulation of general processes,'' {\em Physical Review Letters}, vol.~125,
  no.~1, p.~010501, 2020.

\bibitem{j14}
X.~Yuan, S.~Endo, Q.~Zhao, Y.~Li, and S.~C. Benjamin, ``Theory of variational
  quantum simulation,'' {\em Quantum}, vol.~3, p.~191, 2019.

\bibitem{su}
J.~Schlienz and G.~Mahler, ``Description of entanglement,'' {\em Physical
  Review A}, vol.~52, no.~6, p.~4396, 1995.

\bibitem{r25}
R.~F. Werner, ``Quantum states with einstein-podolsky-rosen correlations
  admitting a hidden-variable model,'' {\em Physical Review A}, vol.~40, no.~8,
  p.~4277, 1989.

\bibitem{r26}
D.~Girolami and G.~Adesso, ``Quantum discord for general two-qubit states:
  analytical progress,'' {\em Physical Review A}, vol.~83, no.~5, p.~052108,
  2011.

\end{thebibliography}
\end{document}